\documentclass[footinbib,amsmath,amssymb,superscriptaddress,prl,reprint]{revtex4-1}
\usepackage[utf8]{inputenc}
\usepackage{graphicx}
\usepackage{xcolor}

\usepackage[version=3]{mhchem}
\usepackage{siunitx}

 \usepackage{comment}

\newcommand{\rs}[1]{_{\rm #1}} 

\usepackage{printlen}
\uselengthunit{in}

\begin{document}

\title{Nonmonotonic behavior in the dense assemblies of active colloids}
\author{Natsuda Klongvessa}
\author{F\'elix Ginot}
\altaffiliation[Now in: ]{Fachbereich Physik, Universität Konstanz, Universitätsstrasse 10, 78464 Konstanz, Germany}
\author{Christophe Ybert}
\author{C\'ecile Cottin-Bizonne}
\author{Mathieu Leocmach}
\email{cecile.cottin-bizonne@univ-lyon1.fr}
\affiliation{Université de Lyon, Université Claude Bernard Lyon 1, CNRS, Institut Lumière Matière, F-69622, VILLEURBANNE, France}

\begin{abstract}
We study experimentally a sediment of self-propelled Brownian particles with densities ranging from dilute to ergodic supercooled, to nonergodic glass, to nonergodic polycrystal.
In a compagnon letter, we observe a nonmonotonic response to activity of relaxation of the nonergodic glass state: a dramatic slowdown when particles become weakly self-propelled, followed by a speedup at higher activities. 
Here we map ergodic supercooled states to standard passive glassy physics, provided a monotonic shift of the glass packing fraction and the replacement of the ambient temperature by the effective temperature.
However we show that this mapping fails beyond glass transition. 
This failure is responsible for the nonmonotonic response.
Furthermore, we generalize our finding by examining the dynamical response of an other class of nonergodic systems : polycrystals.
We observe the same nonmonotinic response to activity.
To explain this phenomenon, we measure the size of domains were particles move in the same direction. This size also shows a nonmonotonic response, with small lengths corresponding to slow relaxation. 
This suggests that the failure of the mapping of nonergodic active states to a passive situation is general and is linked to anisotropic relaxation mechanisms specific to active matter.
\end{abstract}

\maketitle


\section{Introduction}
In the last decade, the mesmerizing dynamic patterns of bird flocks has become a rallying sign for a large community of physicists~\cite{Ballerini2008}. The growing field of active matter deals with the statistical physics of self-propelled objects and has deep implications from crowd dynamics, to energy harvesting, to cancer metastasis. Active systems are driven out of equilibrium by energy injected at the level of the individual particle~\cite{Marchetti2013,Bechinger_rmp-2016, Fodor2018}. Despite their intrinsically nonequilibrium nature, effective thermodynamic variables, e.g. an effective temperature, can be defined to map different steady-state behaviors of active systems onto equilibrium concepts: sedimentation-diffusion~\cite{Palacci2010}, phase separation~\cite{Fily2012, Wysocki2014, Digregorio2018}, or crystallization~\cite{Digregorio2018, Briand2018}. However many collective behaviors emerging from self-propelled systems have no equilibrium equivalent: giant density fluctuations~\cite{Deseigne2010}, clustering~\cite{Theur2012,Ginot2018}, travelling polar phase~\cite{Bricard2013} or turbulence~\cite{Nishi2015}.

In this context, the possibility of such a mapping for nonergodic states of matter, where the system can explore only a small part of the phase space, has received little attention. 
A crystal with frozen-in defects is nonergodic. This is all the more obvious in a polycrystal where grain boundaries are pinned by defects. Studies of active crystals have focused on the shift of the phase boundaries~\cite{Digregorio2018} or on the stability of the crystal lattice at densities lower than close packing~\cite{briand2016CrystallizationSelfpropelledHarddiscs, Briand2018}. In the latter case, alignment interactions between particles can result in an ergodic, ever flowing crystal state~\cite{Briand2018}.

The epitome of ergodicity breaking is the glass transition. In the case of a suspension, the dynamics slows down by orders of magnitude upon compression until the system cannot be equilibrated in a reasonable time. Our understanding of glass as a fundamental state of matter, and of the dynamical arrest that leads to it has tremendously progressed in the last decades through theories that directly address its nonergodic nature~\cite{Cavagna2009,CharbonneauReview2017}. Numerical studies of self-propelled systems approaching the glass transition found that despite a quantitative shift of the glass transition line, the qualitative phenomenology of glassiness remained unchanged~\cite{Berthier2017}. By contrast, in a letter accompanying this article, we show experimentally that the response of the nonergodic glass phase to low levels of self-propulsion is nontrivial and displays Deadlock from the Emergence of Active Directionality (DEAD)~\cite{klongvessa2019a}. In the present article, we address directly the failure of mapping to a passive counterpart nonergodic active states: both glassy and polycrystalline.

The first numerical study of glassy systems of self-propelled particles used the Active Brownian Particle (ABP) model, where the particles are submitted both to propulsion forces and temperature-induced Brownian agitation~\cite{Ni2013b}. Colloidal particles that self-propel by self-phoretic mechanisms are well described by ABP model~\cite{Palacci2010,Theur2012,Ginot2015}. An effective temperature can be defined for dilute ABP that takes into account both the temperature of the bath and the characteristics of the propulsion force~\cite{Palacci2010,hagen2011BrownianMotionSelfpropelled}. For simplicity, later models discarded the thermal bath and kept only various implementations of a persistent propulsion force~\cite{Berthier2014,Szamel2015,Flenner2016,Berthier2017,Nandi2018}. Each of these models can be unambiguously assigned an effective temperature.
Depending on the numerical model, a rise of effective temperature can either lead to activity-induced fluidization~\cite{Ni2013b,Berthier2014} or arrest~\cite{Szamel2015,Flenner2016}. 
It was found that the glass transition line shifted in nontrivial ways with the persistence time of the propulsion direction. The influence of activity could thus not be captured by a single parameter such as effective temperature. For example in Ref.~\cite{Berthier2017} glass transition shifts to higher densities with increasing persistence time if the effective temperature is low, but to lower densities if the effective temperature is high. Indeed, \citet{Nandi2018} have recently shown analytically that the monotonicity of the glass transition shift depends on the microscopic details of the activity.

The focus of the above-cited numerical and analytical studies was on the position of the glass transition line, inferred by an approach from the ergodic supercooled state, and not on the nonergodic state beyond this line.
This is precisely beyond this ergodic to nonergodic line that we find experimentally a nontrivial phenomenology that cannot be mapped to a passive counterpart.
In the accompanying letter~\cite{klongvessa2019a}, we study the influence of self-propulsion on a sediment of Brownian particles, in order to access states on both sides of the nonergodic glass transition. We show that the relaxation of the supercooled liquid speeds up with activity, whereas the nonergodic glass displays a slowdown upon introduction of low levels of activity, but a speedup at higher activity levels.
In the present article, we investigate this phenomenon through the lens of different observable and additional experimental data.
We then perform careful measurements of effective temperature and density in order to map the ergodic supercooled regime at various activity levels onto the passive case. We observe a failure of this mapping beyond the glass transition. We then generalize this observation to polycrystals, an other class of nonergodic systems. By investigating the microscopic relaxation mechanisms inside a polycrystal, we show that the lengthscale relevant to relaxation is not the same as in a passive system and is linked to collective motion.
Finally, we discuss our results in the framework of a competition between cooperative and collective relaxations.


\section{Experimental set-up}
We make Janus particles starting from gold particles (Bio-Rad \#1652264) of diameter \SI{1.6}{\micro\metre} (polydispersity 10\%) that we half coat with \SI{20}{\nano\metre} platinum following the method in Ref.~\cite{Howse2007}.
After purification and sorting, the Janus particles dispersed in deionized water are put into a well (Falcon \#353219). Due to their high density $\rho\simeq \SI{11}{\gram\per\cubic\centi\metre}$, particles settle down to the bottom to form a monolayer. We observe their 2D motion from below on a Leica DMI 4000B microscope equipped by an external dark-field lightning ring and a Basler camera (acA2040-90um). Video data are taken at 5 and \SI{20}{\hertz} and analyzed using Trackpy package~\cite{trackpy2016}.

\begin{figure}
\includegraphics{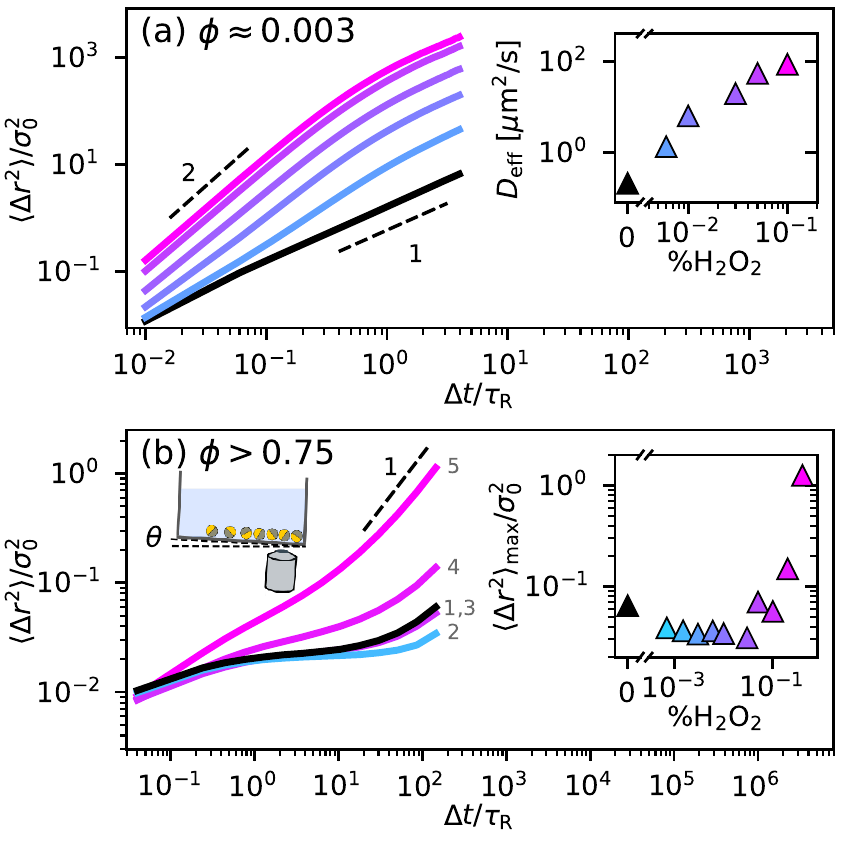}
\caption{Characterization of active colloid motion in dilute (a) and dense (b) regimes at various \ce{H2O2} concentrations increasing from black (without \ce{H2O2}), to cyan to magenta, see respective insets.
(a) Mean square displacement in the dilute regime. 
Dashed lines indicate slopes 1 (diffusive motion) and 2 (ballistic motion). 
Inset: Corresponding effective diffusion coefficient versus \ce{H2O2} concentration, extracted from the long-time MSD.
(b) Mean square displacement in the dense regime. 
The numbers denote the order of increment of \ce{H2O2} concentration.
To increase readability, intermediate concentrations between \#2 and \#3 are not shown, since their curves are almost identical to \#2.
The experimental set-up to obtain the dense regime by tilting the microscope together with the sample is sketched. 
Note that the set-up is not tilted in (a). 
Inset: Values of MSD at the longest lag time.
}
\label{fig:msd_deff}
\end{figure}

Since the colloidal particles are charged and the ionic force of the solvent is low, electrostatic repulsion prevents direct contact between particles. We estimate an effective diameter of the particles to $\sigma_0=\SI{2.2}{\micro\metre}$ from the position of the first peak of the radial pair correlation function in a dense passive regime. This allows us to define the area fraction as $\phi=4\varrho/(\pi \sigma_0^2)$ , where $\varrho$ is the number density. However $\sigma_0/2$ is larger than the hydrodynamic radius of the particles $R_\mathrm{H}$. From the translational diffusion coefficient in dilute conditions, we estimate  $R_\mathrm{H} \approx \SI{0.94}{\micro\metre}$, which corresponds to 
a Brownian translational time $\tau_\mathrm{T}=(3\pi\eta R_\mathrm{H}^3)/(2k_\mathrm{B}T_0) \approx \SI{0.9}{\second}$ and a Brownian rotational time $\tau_\mathrm{R} = (8\pi\eta R_\mathrm{H}^3)/(k_\mathrm{B}T_0) \approx \SI{5}{\second}$, where $T_0$ is the bath temperature. 

Self-propulsion is made possible in dilute hydrogen peroxide (\ce{H2O2}, Merck Millipore, \#1072090250) solutions by a combination of electrophoresis and diffusiophoresis effects~\cite{Paxton2004,Brown2014}. The two halves of the catalytic splitting of \ce{H2O2} occur respectively on each side of the particle, causing self-phoretic effects that drive the particle forward. In dilute regime, the mean square displacement (MSD, see Fig.~\ref{fig:msd_deff}a) displays ballistic motion at short times and diffusive motion at long times due to rotational diffusion.
The rotational diffusion time is practically independent on \ce{H2O2} concentration, and approximately equal to $\tau_\mathrm{R}$.
By contrast, the propulsion velocity increases monotonically with \ce{H2O2} concentration, up to \SI{10}{\micro\metre\per\second} at 0.1 v/v \% concentration. We extract the effective diffusion coefficient, $D\rs{eff}$, of this persistent random motion by fitting the long-time scale MSD and show in the inset that it increases monotonically with \ce{H2O2} concentration.

Due to the large volume of solvent above the monolayer, we find that the effects of activity are stable in time over the course of several hours. In particular, purely diffusive motion could be recovered only several days after the last \ce{H2O2} introduction. That is why we always wash several times our particles with milliQ water before starting a series of experiments and always increase step by step \ce{H2O2} concentration from that clean state. At each step, acquisition is started 30 minutes after \ce{H2O2} introduction to allow a steady state to be reached.

In-plane sedimentation is obtained by tilting the whole set-up with a small angle $\theta \approx \ang{0.1}$, see the sketch of the set-up in Fig.~\ref{fig:msd_deff}b. The monolayer of particles is thus under an in-plane gravity $g\sin\theta\approx \SI{2e-2}{m\per\square\second}$.
The sample is mounted on a motorized XY translation stage (SCAN IM 130x85) that we program for systematic observations at different heights of the sediment with positioning repeatability below \SI{1}{\micro\meter} while minimizing in-plane acceleration.

In the following we will first focus on the densest part of the sediment as we progressively increase activity to characterize its dynamics, effective temperature and density. Then, we will take a broader look at the whole sediment to characterize the system at all densities. In particular, we will characterize dynamics on both sides of the glass transition. Finally, we will focus on individual particles motion of the nonergodic region and discuss how activity affects the relaxation mechanisms.

\section{Dense active behavior}

Here, we recall briefly the nontrivial response of glass dynamics to activity that we observed in a companion letter~\cite{klongvessa2019a}. For this, we use a different series of experiments where we focus on a large region at constant height in the sediment. In the passive case, the packing fraction in this region is $\phi\approx 0.75$.

The black curve in Fig.~\ref{fig:msd_deff}b shows the MSD before any \ce{H2O2} addition (curve \#1). The plateau is typical of glassy behavior and indicates that each particle is trapped by its neighbors. At long times, the system exits the plateau hinting that the particles manage to diffuse away from their original positions~\cite{Weeks2000}.

As we introduce a small amount of \ce{H2O2}, the plateau gets longer, as shown on curve \#2 in Fig.~\ref{fig:msd_deff}b. This surprisingly indicates that the system is less mobile when each particle is weakly self-propelled.
However, when \ce{H2O2} is further increased, we recover a mobility equivalent to the passive case (curve \#3). At even higher concentrations, the system has more and more mobility. The plateau becomes shorter (curve \#4) and finally disappear (curve \#5) where we can observe an effective diffusion motion at long time scale.

In the inset of Fig.~\ref{fig:msd_deff}b, this nonmonotonic behavior is quantified at more values of \ce{H2O2} concentration by the value of MSD at the maximum lag time. Indeed, at concentrations of \ce{H2O2} lower than $\approx 0.04~\%$ the system is more arrested than in the passive case, but diffusion becomes more effective over $\approx 0.10~\%$. 
In the companion letter~\cite{klongvessa2019a}, we propose that such a nonmonotonic response in the glass state is due to two contradictory effects of activity: (i) providing extra energy to the system that helps breaking cages (ii) directional space exploration that is inefficient to explore a cage.
We call the resulting nonmonotonic behaviour Deadlock from the Emergence of Active Directionality (DEAD).

Here we adopt a broader view on nonergodicity by considering both glass and polycrystal. But first, we need to map properly ergodic active states to the passive situation, and test how this mapping fares in nonergodic situations.

\section{ABP framework}

\begin{figure}
\includegraphics{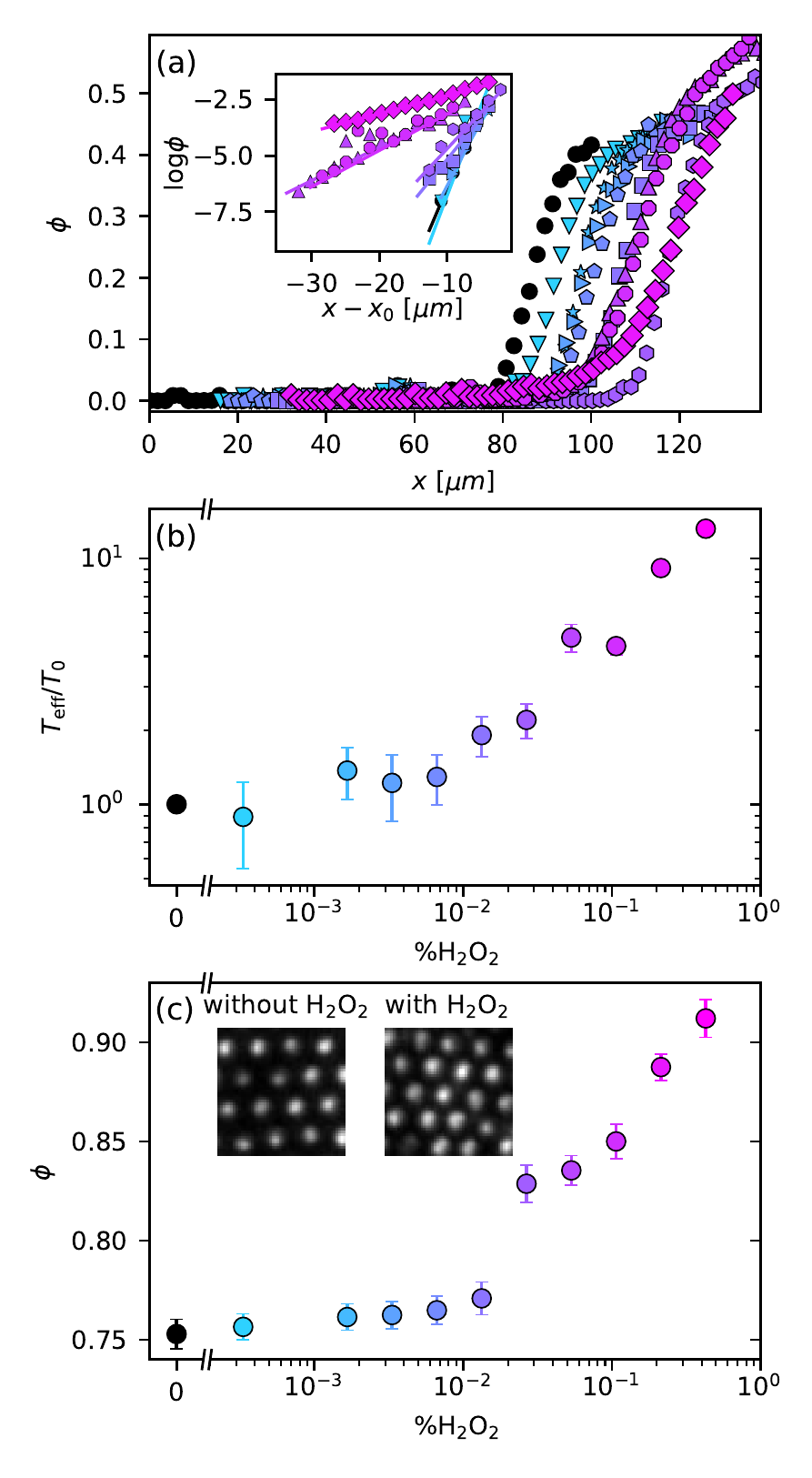}
\caption{(a) Density profile, $\phi(x)$, at the top of the sediment for various \ce{H2O2} concentration color-coded as in Fig.~\ref{fig:msd_deff}b. $x$ is the coordinate in the direction of $g\sin\theta$. (inset) Linear fit of $\log\phi$ in order to obtain the ratio between the effective temperature and the Brownian temperature, $T\rs{eff}/T_0$. The abscissa is shifted by $x_0$, the  position where the profile has the maximum slope. The uncertainty is higher for the passive and low activity cases, where the density profile is sharp and the dilute region is very limited. (b) Calibration of $T\rs{eff}/T_0$ versus \ce{H2O2} concentration. The error bar comes from the uncertainty on the slope measurement. (c) Area density function versus \ce{H2O2} concentration. (inset) Details of experimental images showing the same compaction.}
\label{fig:Teff}
\end{figure}

Experimentally, our particles are submitted to both Brownian and active motions, and are well described by the Active Brownian Particle model, where the 2D persistence time is fixed by 3D Brownian rotational diffusion and thus practically constant $\tau_\mathrm{P} = \tau_\mathrm{R}/2$~\cite{hagen2011BrownianMotionSelfpropelled}, as confirmed in dilute conditions. In a previous work, some of us have shown that the behaviour of the same particles in locally dense clusters can be quantitatively explained without density dependence of the persistence time or alignment interaction between particles~\cite{Ginot2018}. This is why we consider $\tau_\mathrm{P}$ constant throughout the sediment and independent of \ce{H2O2} concentration.

What is changing with \ce{H2O2} concentration is the propulsion force $F_\mathrm{P}$.
%
We cannot measure directly $F_\mathrm{P}$ in a dense sediment. However we have access to an effective temperature ($T\rs{eff}$) measured from the dilute limit of the sedimentation profile, as described in \cite{Ginot2015}.
From the sedimentation experiment on passive colloids~\cite{Perrin1909}, the competition between diffusive motion and gravity $g$ results in a density profile that has the Boltzmann form at low enough densities: $\phi(x) \sim \exp[\Delta m g\sin\theta x/\mu D_0]$, where $\Delta m$ is the buoyant mass, $x$ the coordinate in the direction of gravity, $\mu = 6\pi\eta R_\mathrm{H}$ the mobility and $D_0=k\rs{B}T_0/\mu$ is the diffusion coefficient. Following Refs~\cite{Tailleur2009,Palacci2010}, in the case of self-propelled particles $D_0$ can be replaced by the long time effective diffusion coefficient $D_\mathrm{eff}(\phi \rightarrow 0)$. Following~\cite{Palacci2010,hagen2011BrownianMotionSelfpropelled} we use the case of spherical particles undergoing both Brownian and self-propelled motions in 2D but with two degrees of rotational freedom:
\begin{equation}
    D_\mathrm{eff}(\phi\rightarrow 0) = D_0 + \frac{1}{6} \left(\frac{F_\mathrm{P}}{\mu}\right)^2\tau_\mathrm{R}.
    \label{eq:Deff}
\end{equation}

Equivalently $T_0$ can be replaced by an effective temperature such that $k\rs{B}T\rs{eff}\equiv \mu D_\mathrm{eff}(\phi \rightarrow 0)$. This amounts to viewing a dilute active system as ``hot colloids'' with an effective temperature~\cite{Palacci2010}.
\begin{equation}
    \frac{T_\mathrm{eff}}{T_0} = \frac{D_\mathrm{eff}}{D_0} = 1 + \frac{2}{9}\left(\frac{F_\mathrm{P}R_\mathrm{H}}{k_\mathrm{B}T_0}\right)^2.
    \label{eq:TeffPeH}
\end{equation}
This equation relates directly the propulsion force anywhere in the sediment to measurements performed in the dilute limit of the density profile. In the following we will use $T_\mathrm{eff}$ to characterize activity levels throughout the sediment, including the dense regime.

For each \ce{H2O2} concentration, to characterize the very same experimental conditions, we acquire data at two fixed locations: one near the bottom (dense regime, typically $\phi>0.75$, see Fig.~\ref{fig:msd_deff}b) and then immediately one at the top of the sediment ($\phi$ between 0.4 and 0.5 at the bottom of the image and vanishing quickly with altitude). From the latter, we obtain the density profiles shown in Fig.~\ref{fig:Teff}a. Low density regime indeed displays an exponential dependence (inset) from which we extract $T\rs{eff}/T_0$. The effective temperature dependence on \ce{H2O2} concentration is monotonic, as shown in Fig.~\ref{fig:Teff}b. Using (Eq.~\ref{eq:TeffPeH}), we deduce that $F_\mathrm{P}$ is also monotonic.


\section{Density controlled investigation}

\begin{figure*}
    \centering
    \includegraphics{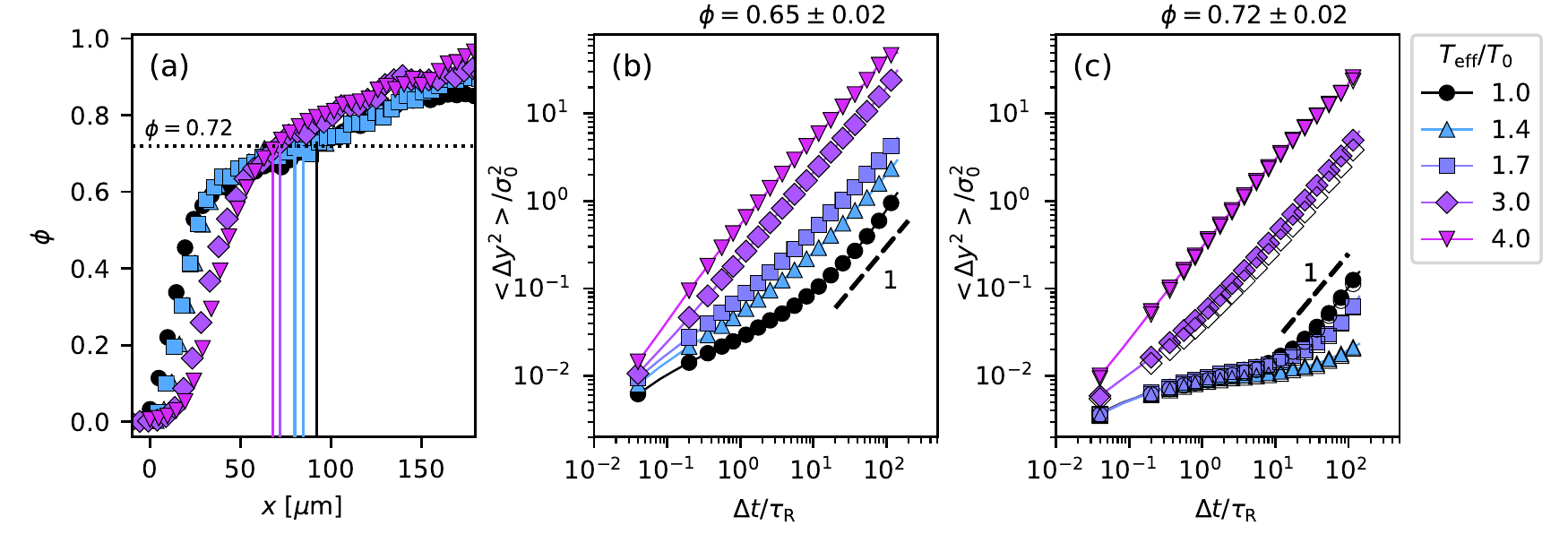}
    \caption{(a) Full density profile $\phi(x)$ comparing between various activity levels. The solid lines illustrate how we match density by moving along altitude $x$.
    MSD along $y$ direction at two fixed densities, in the ergodic (b) and nonergodic (c) phase, and various activity levels. Crystalline particles are excluded in the computation of the MSD in both panels (full symbols). 
    The empty symbols in (c) are the MSD of the crystalline particles (mostly superimposed on full symbols). 
    The dashed lines emphasize diffusive motion. }
    \label{fig:Ft_fix_density}
\end{figure*}

In Fig.~\ref{fig:Teff}a we notice that the sedimentation profiles shift toward larger $x$ with increasing \ce{H2O2} concentration.
This compaction of the sediment is confirmed by measuring the average density in the bottom part, see Fig.~\ref{fig:Teff}c. The particle density rises by about 15\% from the passive to the highest activity and we observe that the inter-particle distance also becomes smaller (inset). 
Compaction could be due to purely chemical effects caused by the increase of \ce{H2O2} concentration, or could be a general feature of self-propelled particles confined by an external potential. More probably it is a combination between these two factors, as we observe that a sediment of uncoated gold particles does compact with \ce{H2O2} concentration but by only 4\%. 
Such compaction is consistent with the effective attraction some of us observed in the same system at lower densities~\cite{Ginot2015}. Attractive interactions can significantly alter the glass transition scenario of passive systems in isochoric conditions~\cite{Berthier2011e} but has no influence if the system is able to adjust its volume (isobaric conditions)~\cite{dell2015MicroscopicTheoryRole}. Here our system is not isochoric but confined by gravity, therefore, if the effect of activity is solely an effective interparticle attraction, we expect a trivial mapping of dynamics onto the purely repulsive passive system.

In any case, glassy phenomenology is extremely sensitive to density variations and we have to control for this parameter before reaching to any conclusion. We thus perform another set of experiments where we observe the whole density profile. As shown on Fig.~\ref{fig:Ft_fix_density}a, for most values of $\phi$, we can select activity-by-activity the position in the density profile that corresponds to the density $\phi$. A thin slice orthogonal to gravity centered on this position has thus an average density of $\phi$. We can thus follow the density $\phi$ at all activities and work at constant density.

In the companion letter, we have identified the glass transition packing fraction in the passive case $\phi_\mathrm{g}(T_0) \approx 0.67$.
In Fig.~\ref{fig:Ft_fix_density} we show MSD for various activities but at two fixed densities, on both sides of $\phi_\mathrm{g}(T_0)$. 
They show striking contrast that can be directly interpreted in terms of cage size.

At $\phi= 0.65\pm 0.02 < \phi_\mathrm{g}(T_0)$ (Fig.~\ref{fig:Ft_fix_density}b), the shape of the MSD evolves monotonically with $T\rs{eff}/T_0$. The passive case displays a subdiffusive plateau, which level increases with activity until total disappearance at the two highest activities. The increase in plateau height from the passive case to $T\rs{eff}/T_0 = 1.4$ and $1.7$ indicates wider cages.

At $\phi= 0.72\pm 0.02$, the dynamics of the system shows stark differences. The height of the plateau in the MSD (Fig.~\ref{fig:Ft_fix_density}c) does not depend on activity at low levels, hinting at a constant cage size ($\approx 0.3\sigma_0$). However, the exit of the plateau does depend on activity in a nonmonotonic way. Activity $T\rs{eff}/T_0 = 1.4$ exits the plateau later than the passive case. The next activity exits earlier than $1.4$ but still later than the passive case. The two last activities show no plateau. We thus recover the nonmonotonic behavior, even at constant density.

At $\phi= 0.72\pm 0.02$, weak self-propulsion is not enough to enlarge the accessible area. It reveals that each particle faces steep energy barriers. Particles are already as close as they can be. Since their interaction potential is steep at short distances, the extra energy afforded by $T\rs{eff}/T_0 <2$ cannot push the particles significantly closer, which shows on the constant plateau level of MSD.
By contrast at $\phi= 0.65\pm 0.02$ the particles are relatively further apart, feeling a softer confinement. Therefore, even weak self-propulsion can push against these barriers and enlarge the accessible area, which shows on the increasing plateau level of MSD.

\section{Effect of local structure}

The polydispersity of the particles and the presence of doublet or triplet aggregations are not sufficient to completely prevent crystal nucleation at high enough density. 
We quantify the degree of local ordering using the hexatic order parameter~\cite{David1979}:
\begin{equation}
\psi_{6,i} = \frac{1}{6}\sum_{j\in n_i} \exp(6i\theta_{i,j})
\end{equation}
where $n_i$ is the set of 6-nearest neighbors of the particle $i$ and $\theta_{i,j}$ is the angle of the vector between particle $i$ and particle $j$ with respect to the reference frame. Particles with $|\psi_{6,i}|>0.8$ are considered crystalline. In the passive case, the ratio of crystalline particles raises from 20\% at $\phi=0.65\pm0.02$ to 40\% at $\phi=0.72\pm0.02$. We verify that due to the strong gravity confinement in our system, a crystal nucleus has the same density as its amorphous surroundings. Therefore each slice has a well-defined $\phi$.

Even when local order and density are decoupled, the presence of local order can have a large influence on the dynamics of glassy systems~\cite{tanaka2010critical}. However, here we find little difference between the MSD of the crystalline particles and the non-crystalline particles of the same slice at the same activity, see empty symbols on Fig.~\ref{fig:Ft_fix_density}c.
Nevertheless, in order to focus on glassy dynamics, we first exclude crystalline particles from our analysis, as well as slices that contains more than 50\% crystalline particles. Later on, we will analyze further the consequences of this choice and the dynamics in polycrystalline slices.

\section{Mapping active glassy behaviour to equilibrium}

\begin{figure*}
\includegraphics{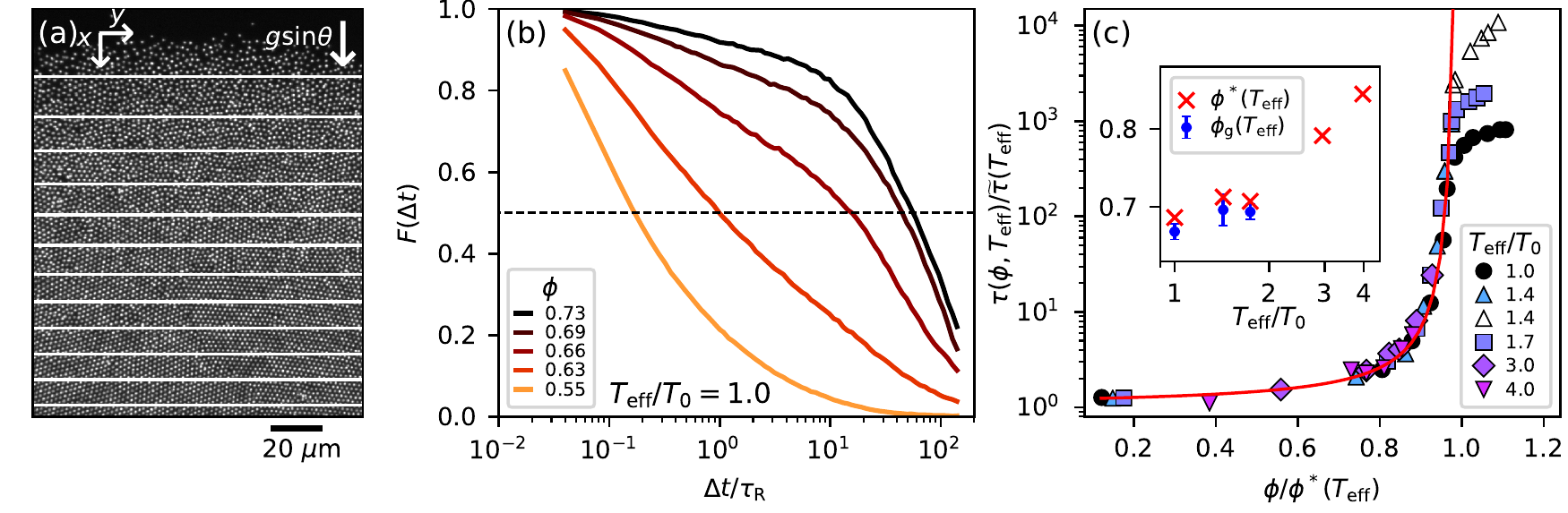}
\caption{(a) Experimental image of the sediment showing the slicing to get access to different densities.
(b) Overlap function $F(\Delta t)$ in the passive case and  at various densities. The dashed line at 0.5 is the threshold where the relaxation time $\tau$ is defined. 
(c) Collapse of density dependence of relaxation time on Eq.~\ref{eq:collapse} (red curve). Beyond glass transition collapse is lost and saturation level follows a nonmonotonic trend with activity.
Open triangles are obtained by extrapolation of $F(\Delta t)$.
Inset: Ideal and operational glass transition packing fractions, $\phi^*$ and  $\phi\rs{g}$ respectively, function of activity.
}
\label{fig:slice}
\end{figure*}

In order to characterize the dependence of glass transition on activity, we perform systematic analysis of the dynamics function of density and activity.
We divide the sediment into thin slices perpendicular to the gravitational gradient (see Fig.~\ref{fig:slice}a). We set the width of each slice so that every slice contains approximately the same number of particles ($1000\pm 100$ particles per slice). We can compute all static and dynamic quantities function of the density, parameterized by the altitude $x$. Crystalline particles are excluded.

To characterize the relaxation dynamics, we define a microscopic overlap function $w_i(t_0,\Delta t) = \Theta(a -\|\vec{r}_i(t_0+\Delta t) - \vec{r}_i(t_0)\|)$ that indicates whether particle $i$ has not moved further than $a = 0.3\sigma_0$ between times $t_0$ and $t_0+\Delta t$. The value of the threshold distance $a$ corresponds to the height of the plateau of the MSD and thus to the cage size. Here $\Theta$ is the Heaviside step function.
In each slice, we compute the overlap function~\cite{Flenner2011}, $F(\Delta t)$, which tells us the ratio of particles that have not moved:
\begin{equation}
F(\Delta t)=\left\langle \frac{1}{N} \sum_{i=1}^N w_i(t_0,\Delta t) \right\rangle_{t_0}.
\end{equation}

Fig.~\ref{fig:slice}b shows $F(\Delta t)$ at various densities of the passive sediment. At high densities, we observe a two-step relaxation typical of glassy dynamics. 
We note that contrary to systems with a steep repulsive potential, here the height of the plateau depends on density. 
The plateau completely disappears at the lowest density, and $F(\Delta t)$ relaxes in a single exponential step indicating nonglassy behavior.
We define the relaxation time $\tau$ when half of the particles have already relaxed, i.e., $F(\tau) = 0.5$ (horizontal dashed line).

For each activity, the density dependence of $\tau$ is well fitted by the expression
\begin{equation}
    \frac{\tau(\phi,T_\mathrm{eff})}{\widetilde{\tau}(T_\mathrm{eff})} = \exp\left[\frac{A}{\left(\phi^*(T_\mathrm{eff})/\phi\right) -1}\right],
    \label{eq:collapse}
\end{equation}
where $A\approx 0.19$ is independent of activity, whereas $\widetilde{\tau}(T_\mathrm{eff})$ and $\phi^*(T_\mathrm{eff})$ are activity-dependent parameters, respectively the relaxation time in the dilute limit and the packing fraction at which the fit diverges, often called the ideal glass transition packing fraction.

In Fig.~\ref{fig:slice}c we observe the collapse of all activities onto (Eq.~\ref{eq:collapse}). However, this collapse does not hold beyond the glass transition, where $\tau$ only depends weakly on $\phi$. This trend contradicts the usual picture of glass transition where the relaxation time should diverge. However, our phenomenology is consistent with what \citet{philippeGlassTransitionSoft2018} have observed in a large variety of passive systems made of soft particles. We define the operational glass transition density $\phi_\mathrm{g}(T\rs{eff})$, as the packing fraction at which the system becomes nonergodic. The inset of Fig.~\ref{fig:slice}c shows our estimate of $\phi_\mathrm{g}$ as where the data departs from the master curve for each activity. For $T\rs{eff}/T_0 = 3.0$ and $4.0$, $\phi\rs{g}(T\rs{eff})$ cannot be defined because the ratio of crystalline particles reaches 50\% without deviation from the master curve. 

The collapse of the supercooled regimes in Fig.~\ref{fig:slice}c indicates that $\phi^*$ and $\widetilde{\tau}$ are enough to describe the physics of glass transition below $\phi_\mathrm{g}$.
However, above $\phi_\mathrm{g}$, in the nonergodic regime, $\tau/\widetilde{\tau}$ saturates. This saturation value is different at each activity. It suggests that $\widetilde{\tau}(T_\mathrm{eff})$ and thus an effective temperature is not enough to describe the effects of self-propulsion on the nonergodic glass. Moreover, this saturation value gives a hint of the nonmonotonic DEAD behavior: an order of magnitude jump between the passive case and the first nonzero activity, and then a decrease with increasing activity. We are thus confident that the DEAD phenomenology originates directly from the particle self-propulsion and is neither a pure effect of (attraction induced) compaction nor reducible to an increase in (effective) temperature. 

We have cornered the nonmonotonic response to activity beyond ergodicity breaking. Glass is a nonergodic state of matter, but so is a defective crystal or a polycrystal with quenched disorder. Comparing Fig.~\ref{fig:Ft_fix_density}b and c, we have already noted that particles in crystal nuclei display the DEAD behavior. In the following, we will analyze the dynamics of fully polycrystalline slices to look for a similar nonmonotonic behaviour.

\section{Relaxation of active polycrystal}

\begin{figure}
    \centering
    \includegraphics{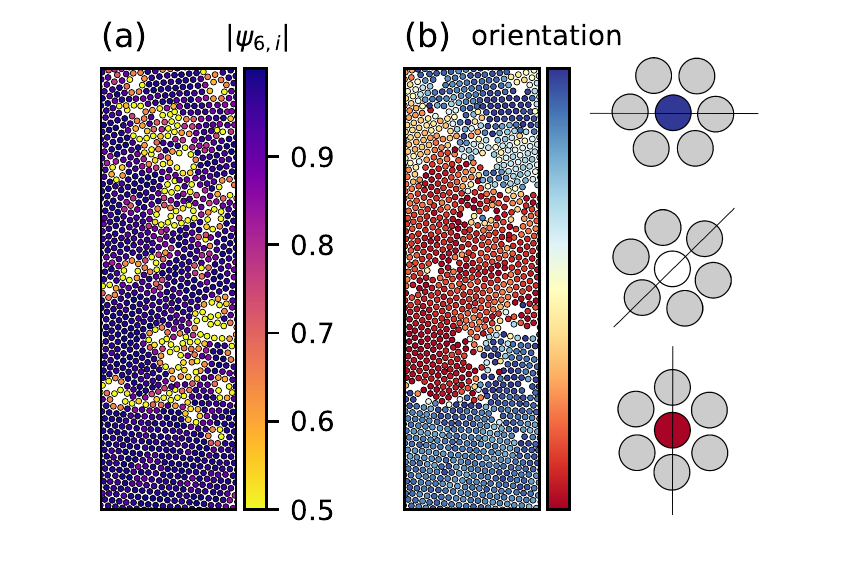}
    \caption{Maps of hexatic structure parameter ($\psi_{6,i}$) of the passive case at $\phi = 0.85 \pm 0.03$: (a) modulus and (b) orientation. Sketches on the right shows which orientation corresponds to which color. The white areas are from sample artifacts and tracking errors. The particles position is exactly the same as in Fig.~\ref{fig:orientation}a,b for the passive case. }
    \label{fig:psi6_map}
\end{figure}

Here, we explore the microscopic details of the relaxation mechanism in a polycrystalline slice width $\approx 60\sigma_0$. The packing fraction is approximately uniform with $\phi = 0.85 \pm 0.03$.
At this high density, the system is highly ordered and 80\% of particles are crystalline, as shown in the map of $|\psi_{6,i}|$ (see Fig.~\ref{fig:psi6_map}a). Following Refs~\cite{Bernard2011,Digregorio2018}, we consider the projection of the phase of $\psi_{6,i}$ as shown in Fig.~\ref{fig:psi6_map}b. We can thus clearly distinguish crystalline domains of consistent orientation separated by sharp grain boundaries where sample impurities concentrates (low $|\psi_{6,i}|$). This slice is indeed polycrystalline and not hexatic.
As we increase activity, there is no obvious difference between the passive and low activities in terms of ordering. Furthermore, the grain boundaries, pinned by sample impurities, remain stable. The lost of ordering can be noticed only at high enough $T\rs{eff}/T_0$. Here the percentage of crystalline particles drops from 80\% in the passive case to 75\% for $T\rs{eff}/T_0 = 3.0$ and to 68\% for $T\rs{eff}/T_0 = 4.0$. 

\begin{figure}
    \centering
    \includegraphics{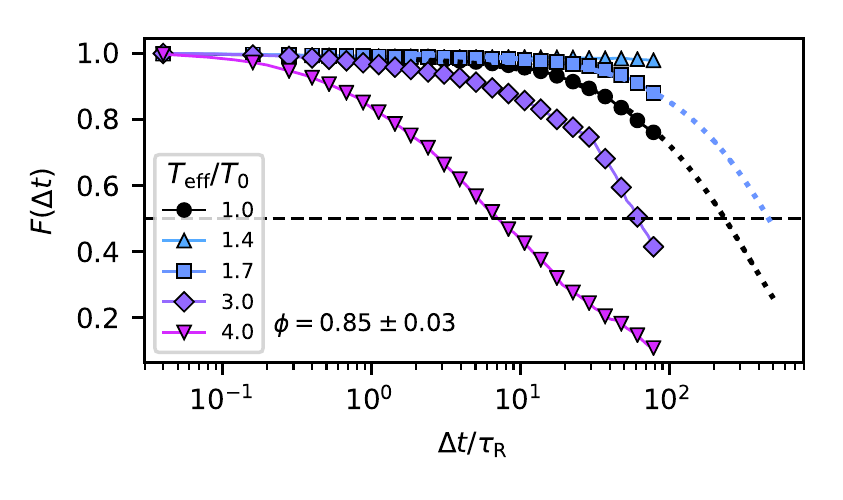}
    \caption{Overlap function in a polycrystalline slice. The threshold where the relaxation time is defined is at $F = 0.5$ (dashed line). The dotted lines at $T\rs{eff}/T_0 = 1.0$ and $1.7$ are the extrapolation of a stretched exponential fit of $F(\Delta t)$ to obtain $\tau$. For $T\rs{eff}/T_0 = 1.4$, $F(\Delta t)$ has not exited the plateau and the extrapolation is not applicable.}
    \label{fig:ft_den80}
\end{figure}

Fig.~\ref{fig:ft_den80} shows the overlap function, $F(\Delta t)$, in this slice at various activity levels. Both crystalline and noncrystalline particles are taken into account. At this density, $F(\Delta t)$ of the passive case and the two lowest activities have not relaxed to the threshold 0.5 within our maximum experimental time. Nevertheless, we can clearly observe the delay in the exit of the plateau. This delay does respond nonmonotonically to activity, in a very similar way to Fig.~\ref{fig:msd_deff}b and c. 
This confirms that such a nonglassy system actually displays DEAD phenomenology.

\begin{figure}
    \centering
    \includegraphics{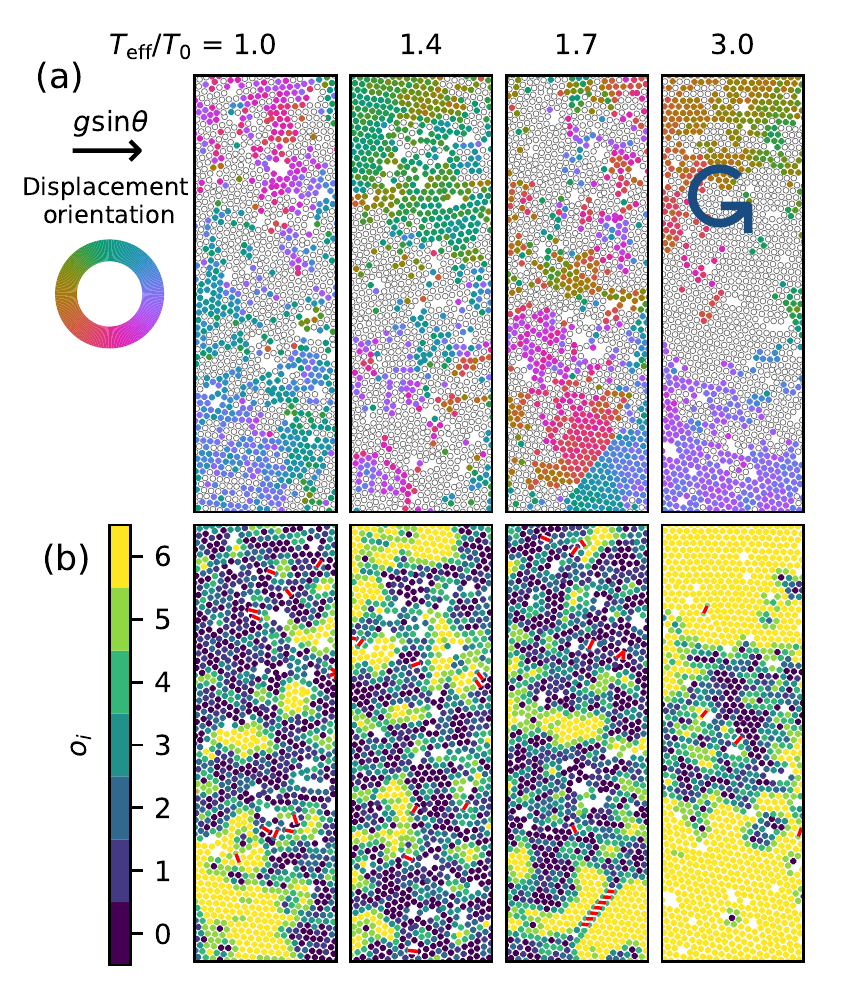}
\caption{
    (a) Orientation of displacement between two frames such that $\Delta t=32\tau_\mathrm{R}$, at various $T_\mathrm{eff}/T_0$ and fixed $\phi=0.85\pm 0.03$.
    Orientations are indicated by colors. 
    The slowest half of particles are shown as empty circles.
    The circle arrow in the last panel highlights the vortex collective motion.
    The white areas are from sample artifacts and tracking errors.
    (b) Directional correlation map that displays for each particle $i$ the number $o_i$ of its six neighbors that have the same orientation of displacement as $i$. The red lines represent broken bonds during $\Delta t$. 
    }
    \label{fig:orientation}
\end{figure}


In order to probe how the system relaxes, we look into the orientation of particle displacement.
To compute displacements $\vec{u}_i$, we focus on the time interval $\Delta t = 32\tau\rs{R}$, which corresponds to $F(\Delta t)$ exiting from the plateau in the passive case (see Fig.~\ref{fig:ft_den80}).
Fig.~\ref{fig:orientation}a spatially maps the orientation of the displacements at different activity levels. 
To highlight large displacements, only the 50\% faster particles are colored according to the orientation of their displacement, while the slower particles are displayed by empty circles.
This representation highlights spatial correlations of the orientations. 
In the fast domains, particles tend to have almost the same direction as their neighbors, and this is true for all activities. 
Furthermore, the boundaries between the domains seem sharper at higher activities. 
We are able to observe shear zones where two zones of opposite orientation slide past each other (3rd panel), and vortices where the particles rotate around a relatively immobile center (4th panel). 
The position, shape and size of these rearrangements bear little correlation with the crystalline grains and grain boundaries identified in Fig.~\ref{fig:psi6_map}.

Next, we characterize further the spatial correlation of orientation displacement. For each particle $i$, we count the number of its neighbors $j$ that have moved to the same direction after $\Delta t$:
\begin{equation}
    o_i = \sum_j \Theta\left(\frac{\vec{u}_i \cdot \vec{u}_j}{|\vec{u}_i||\vec{u}_j|} - 0.5\right),
\end{equation}
where $\Theta$ is the Heaviside step function. Fig.~\ref{fig:orientation}b shows the map of $o_i$ for the same snapshot as in Fig.~\ref{fig:orientation}a 
Although the value of the orientation is lost in this representation, we can clearly observe its spatial correlation. We observe that the fast domains in Fig.~\ref{fig:orientation}a roughly correspond to highly oriented domains in Fig.~\ref{fig:orientation}b.
This hints toward relaxation processes where neighboring particles move together in the same direction. Such collective motions are characteristic of active matter and have been observed from dilute~\cite{Bricard2013} to dense crystalline systems~\cite{Briand2018} provided explicit alignment interactions. 
However here oriented domains are present even in the passive case. This proves that the mechanisms (e.g. dislocation, defect, or grain boundary motion) that makes directed motion emerge from microscopically isotropic motion are already present in the passive polycrystal. Again, no explicit alignment interaction are needed to induce collectively directed behaviour.

To characterize how collective relaxation modes affect the structure of the system, we look for bonds broken over $\Delta t$.
A bond between particle $i$ and particle $j$ at time $t_0$ is defined if (i) particle $j$ is one of the 6-nearest neighbors of $i$ and vice versa, (ii) the distance $r_{ij}$ is shorter than $1.5\sigma_0$. A bond is broken between $t_0$ and $t_0+\Delta t$ if (i) it belongs to the bond network at $t_0$, (ii) it does not belong to the bond network at $t_0+\Delta t$, (iii) both particles $i$ and $j$ are tracked at $t_0+\Delta t$. The broken bonds are presented by red lines in Fig.~\ref{fig:orientation}b. There are very few broken bonds during the relaxation except in shear zones (panel 3). 
It means that at high activity particles move in a correlated manner, such that relative positions between neighbors almost do not relax, despite fast relaxation of absolute positions.

\begin{figure*}
    \centering
    \includegraphics{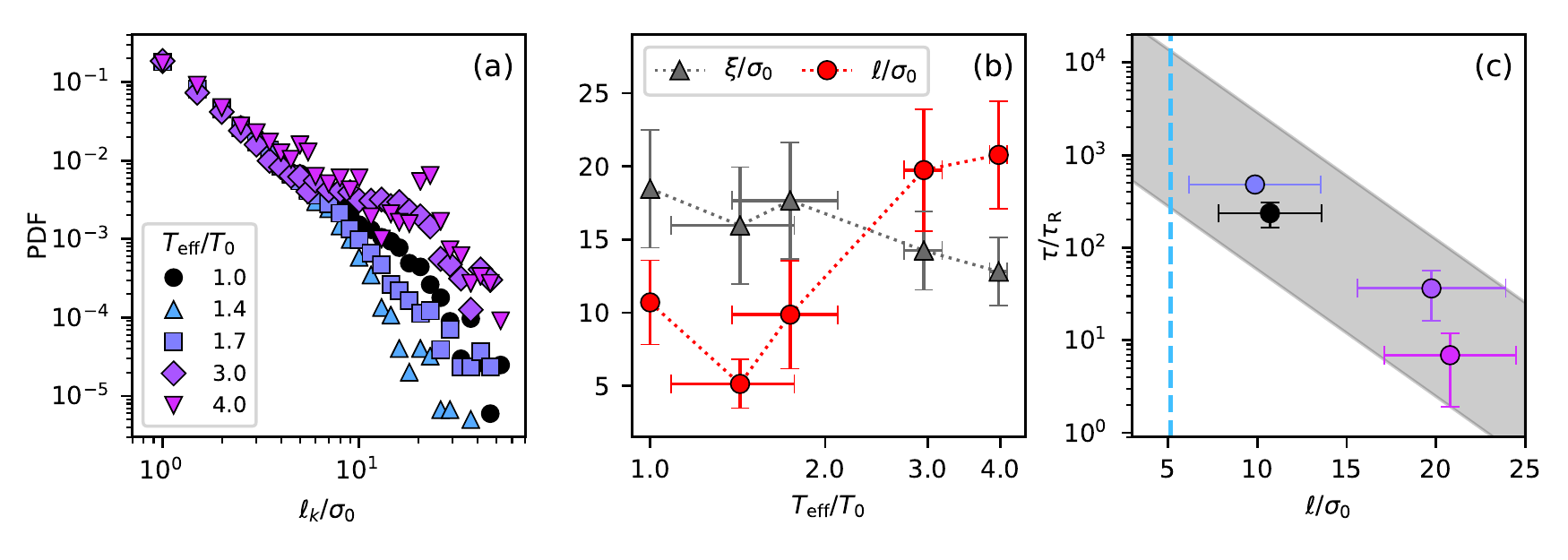}
    \caption{(a) Probability distribution function (PDF) of $\ell_k$ for each domain $k$ at $\phi = 0.85 \pm 0.03$ and various $T\rs{eff}/T_0$. 
    (b) Average size of directional correlation domains $\ell$ (red) and slow domains $\xi$ (gray) at various $T\rs{eff}/T_0$.
    (c) The correlation between the relaxation $\tau$ measured from Fig.~\ref{fig:ft_den80} and size of correlated domains $\ell$. The measurement is done at various $T\rs{eff}/T_0$ color-coded as in (a). The vertical dashed line corresponds to $T\rs{eff}/T_0 = 1.4$ where the relaxation function $F(\Delta t)$ has not yet relaxed within our maximum lag time.}
    \label{fig:ell}
\end{figure*}


In Fig.~\ref{fig:orientation}b we notice qualitatively that state points with faster relaxation seems to have larger correlated  domains. 
To make this observation quantitative, we measure the characteristic size of these domains.
First, we define the domains of the correlated particles ($o_i \geq 4$) and then we define the graph of all particle bonds such that each particle is bonded to its six nearest neighbors. Next, we take the subgraph of the high $o_i$ domains and split it in connected components~\cite{networkX}. This defines correlated domains.
For each correlated domain $k$, we measure its radius of gyration in the $y$ direction (perpendicular to $g\sin\theta$): 
\begin{equation}
    \ell_k = \frac{1}{N_k}\sqrt{\sum_{i\in k} y_i^2 - (\sum_{i\in k} y_i)^2},
    \label{eq:xik}
\end{equation} 
where $N_k$ is the number of particles in cluster $k$. 
The probability distribution function (PDF) of $\ell_k$ of all clusters at all time for various activities are displayed in Fig.~\ref{fig:ell}a. For ($\ell_k/\sigma_0<5)$, the distributions at all activities collapse. Compared to the distribution in the passive case, low activities are deprived of large oriented domains, whereas high activities have an excess probability of large oriented domains. Above $\ell_k\approx10$ the distributions follow a nonmonotonic behavior.

This is confirmed by the characteristic size of the domains, that we define by a weighted average of $\ell_k$ on all clusters at all time: 
\begin{equation}
    \ell = \frac{\sum_{t_0} \sum_k N_k \ell_k}{\sum_{t_0} \sum_k N_k}.
    \label{eq:ell}
\end{equation} 
As shown on Fig.~\ref{fig:ell}b, $\ell$ displays a nonmonotonic evolution with activity consistent with the DEAD behavior: a drop of a factor 2 from the passive case to the lowest activity, and then a progressive increase at higher activities.  
This nonmonotonic response is not captured by the size $\xi$ of the slow domains (defined in the same way as $\ell$, except that the 50\% slower particles are considered instead of the particles where $o_i \geq 4$). $\xi$ is almost constant, with a possible decreasing trend.

From Fig.~\ref{fig:ft_den80}c, we can estimate the relaxation time $\tau$ by extrapolating $F(\Delta t)$. We fit the tail of $F(\Delta t)$ by a stretched exponential $A\exp(-t/\tau_\alpha)^\beta$ and read $\tau$ where the fit crosses the threshold 0.5. 
This procedure is impossible for $T\rs{eff}/T_0 = 1.4$ where $F(\Delta t)$ does not decay significantly.
For all other activities, we can plot $\tau$ function of the length $\ell$. 
Fig.~\ref{fig:ell}c shows that $\ell$ evolves in reverse to what one would expect for a 4-point correlation length in glassy systems.
%
Larger 4-point correlation implies longer relaxation in passive glassy systems~\cite{Cavagna2009}, in active supercooled liquids~\cite{Flenner2016}, and in active crystals with alignement interactions~\cite{briand2016CrystallizationSelfpropelledHarddiscs}.
Here, large $\ell$ corresponds to fast relaxation. Indeed, $\ell$ measures the size of domains with correlated orientation of displacement, associated with collective rearrangements, whereas 4-point correlation measures the size of cooperatively rearranging regions. A large domain moving collectively in the same direction enhances relaxation, whereas a large cooperative region size implies a larger energy barrier and thus longer relaxation. This hints to the existence of relaxation mechanisms specific to self-propelled particles that involve collective directed motion instead of cooperative rearrangements. 

The speedup of the dynamics at high activities can be explained by the rise of collective motion. However the delayed exit from the plateau, characteristic of the DEAD phenomenology occurs when collective motion is still negligible. Therefore, as in the nonergodic glass, our results in the polycrystal point to a drop in efficiency of cooperative rearrangements between the passive case and our lowest activities.

\section{Discussion and conclusion}

We have found experimentally that the approach to glass transition in an active system can be mapped onto the behavior of a passive supercooled liquid of soft colloids. However, we have observed the failure of this mapping beyond the glass transition, characterized by a nonmonotonic response of the relaxation time to an increase of effective temperature. Furthermore we have shown that this phenomenology is not restricted to the amorphous glass, but also observed in polycrystalline regions where grain boundaries are pinned. There, we are able to link the relaxation time to the size of collective motion. We thus evidence that the nonmonotonic behaviour is linked to a drop in the efficiency of cooperative relaxation modes between passive and low-activity cases (Deadlock from the Emergence of Active Directionality) and then the rise of collective motion.

In the companion letter~\cite{klongvessa2019a} we have shown that the initial drop can be at least partly understood in terms of efficiency of cage exploration between Brownian and self-propelled particles. This argument is valid in any nonergodic situation that can be modelled by cage exploration and escape. This is indeed the case of both the glass state and the pinned polycrystalline state that are nonergodic, contrasting to the ergodic liquid. Therefore, we predict that ergodicity breaking is sufficient to preclude mapping to equilibrium of active systems.
However, this simple one-body model does not predict the magnitude of the slowdown, nor the drop in the size of oriented displacements domains. Our work calls for theoretical or numerical investigations in the range of activities where Brownian motion and self-propulsion compete, with a focus on nonergodic states.

\begin{acknowledgments}
The authors thank Ludovic Berthier, Grzegorz Szamel, Chandan Dasgupta and Takeshi Kawasaki for fruitful discussions.
N.K. is supported by PhD scholarship from the doctoral school of Physics and Astrophysics, University of Lyon.
N.K. an M.L. acknowledge funding from CNRS through PICS No 7464. 
M.L. acknowledges support from ANR grant GelBreak ANR-17-CE08-0026.
C.C.B. and C.Y. acknowledge support from ANR grant TunaMix No.
ANR-16-CE30-0028 and from Université de Lyon, within
the program Investissements d’Avenir IDEXLyon (Contract
No. ANR-16-IDEX-0005) operated by ANR.
\end{acknowledgments}

%

\end{document}